\documentclass[aps,pre,reprint,10pt,superscriptaddress,showpacs]{revtex4-1}
\usepackage{amsfonts} 
\usepackage{amsmath}
\usepackage{amssymb}
\usepackage{graphicx}
\usepackage{subfigure}
\usepackage{color}
\newcommand*\xbar[1]{%
  \hbox{%
    \vbox{%
      \hrule height 0.5pt 
      \kern0.5ex
      \hbox{%
        \kern-0.1em
        \ensuremath{#1}%
        \kern-0.1em
      }%
    }%
  }%
} 
\begin{document}
\title{Modulation of kinetic Alfv{\'e}n waves in an intermediate low-beta magnetoplasma}
\author{Debjani Chatterjee}
\email{chatterjee.debjani10@gmail.com}
\author{A. P. Misra}
\email{apmisra@visva-bharati.ac.in; apmisra@gmail.com}
\affiliation{Department of Mathematics, Siksha Bhavana, Visva-Bharati University, Santiniketan-731 235, West Bengal, India}
\begin{abstract}
We study the amplitude modulation of nonlinear kinetic Alfv{\'e}n waves (KAWs) in an intermediate low-beta magnetoplasma. Starting from a set of fluid equations coupled to the Maxwell's equations, we derive a coupled set of nonlinear partial differential equations (PDEs) which govern  the evolution of KAW envelopes in the plasma. The modulational instability (MI) of such KAW envelopes is then studied by a  nonlinear Schr{\"o}dinger (NLS) equation derived from the coupled PDEs.  It is shown that  the KAWs can evolve into bright  envelope solitons, or  can  undergo damping depending on whether  the characteristic ratio  $(\alpha)$ of the Alfv{\'e}n to ion-acoustic (IA) speeds remains above or below a critical value. The parameter $\alpha$ is also found to shift the MI domains around the $k_xk_z$ plane, where $k_x~(k_z)$ is the KAW number perpendicular (parallel) to the external magnetic field.  The growth rate of MI, as well as the frequency shift and the energy transfer rate, are obtained and analyzed.   The   results can be useful for understanding the existence  and formation of  bright and dark envelope solitons, or    damping of KAW envelopes     in space plasmas, e.g.,  interplanetary space, solar winds etc.  
\end{abstract}
\maketitle
\section{Introduction}\label{sec-intro} The kinetic Alfv{\'e}n waves (KAWs) are ubiquitous in space plasmas and   play  an important role in the dissipation of solar wind turbulence \cite{modi2013,sharma2013} and acceleration of  charged particles \cite{cramer2001}. In a general sense, the non-dispersive magnetohydrodynamic (MHD) Alfv{\'e}n waves (AWs) are generated due to the frozen-in field effect by which the charged particles (e.g., electrons) follow the magnetic field lines. However,  when  the perpendicular wavelength of, e.g.,  low-frequency ion oscillations   is of the order of the ion gyroradius (which is much larger than the electron gyroradius), ions are no longer attached to the magnetic field lines, and thus a small difference of the transverse velocities of electrons and ions   creates an electrostatic field parallel to the ambient magnetic field. This electric field, however, modifies the wave dispersion to generate a new kind of mode known as KAW \cite{hasegawa1976}. The latter is also generated  when the  perpendicular (to the magnetic field) component of the AW wave number becomes larger than its parallel component.   So, the KAW is distinctive from the non-dispersive MHD Alfv{\'e}n wave  as not only it has dispersion and Landau dissipation, it is also a  coupling mode (longitudinal-transverse) of  the Alfv{\'e}n wave and the ion-acoustic wave (IAW). We mention that the KAW was first introduced in space physics by Hasegawa  \textit{et. al.} \cite{hasegawa1976}. 
\par
The observations of spacecrafts, like Freja, FAST, Cluster, etc., also indicate the existence of KAWs in intermediate low-$\beta$ space  plasmas \cite{louarn1994,chaston1999,seyler2007}  with $\left({m_e}/{m_i}\right)\ll\beta\ll1$, where $m_e~(m_i)$ is the electron (ion) mass and  $\beta$ is the ratio of the  thermal and  magnetic pressures in the plasma. It has been shown that such low-frequency (compared to the ion-cyclotron frequency) KAWs can evolve into intermediate shocks due to the anomalous dissipation, Alfv{\'e}nic solitons and magnetic kink that are relevant in the solar wind for acceleration of charged particles \cite{liting2001}.  Furthermore, the dispersion properties of KAWs and the formation of localized structures, as well as the onset of turbulence have been investigated owing to their applications in space and astrophysical plasmas \cite{lysak2001,sharma2013,woo2017,sah2010,ahmed2017,bains2014}. 
\par
The modulational instability (MI)  is generally referred to as a first step to study nonlinear processes like the  formation of envelope solitons and their amplitude modulation due to the carrier wave self-interaction, as well as the onset of turbulence in space and astrophysical plasmas \cite{chatterjee2015,chatterjee2016,misra2017,shalini2017,shahmansouri2016,saini2017}. In the process of MI, the transfer of wave energy takes place from high-frequency sidebands to low-frequency ones. Here, the sidebands experience exponential growth due to the pump until some nonlinear effects intervene to restrain the growth. 
 \par
The purpose of the present work is to advance the theory of KAWs, and to investigate the stability and instability criteria for the modulation of KAW packets   in an intermediate low-$\beta$  plasma.  It is   found that the nonlinear KAW can evolve into bright  envelope solitons, or can undergo wave damping due to the anomalous dissipation of KAWs.
\section{Theoritical Formulation}\label{sec-model}
We consider the nonlinear propagation of KAWs in a two-dimensional (2D) intermediate low-$\beta$ $\left({m_e}/{m_i}\ll\beta\ll1 \right)$ plasma composed of  inertialess isothermal electrons and magnetized singly charged isothermal positive  ions. Such plasmas appear, e.g., in the solar wind. In this low-$\beta$ limit, the electron inertia can be neglected (since $m_e/m_i\ll1$), and so one can write down the  electron equation of motion without the term $\sim dv/dt$ [Eq. \eqref{moment-eq-elctron}]. This inertial effect of electrons can be important in the evolution of inertial Alfv{\'e}n waves.  We   assume that the  external uniform magnetic field is    ${\bf B}_0=B_0\hat{z}$ and the electric field in the $xz$ plane, i.e.,  ${\bf E}=E_x\hat{x}+E_z\hat{z}$.  For the excitation of KAWs, the electron gyroradius is much smaller than that of ions, and electrons move along the magnetic field lines. So, the motion  of electrons perpendicular ($x$-component) to the background magnetic field  is negligible. Thus, it is reasonable to consider only the $z$-component of  the equation of motion [Eq. \eqref{moment-eq-elctron}]  for electrons without the Lorentz force (as ${\bf v}\times {\bf B}$ term vanishes along the $z$-axis), and the $z$-component of the equation of motion for magnetized ions with the Lorentz force [Eq. \eqref{moment-eq-ion}]. Furthermore, since in space plasmas, electrons and ions can have thermal motions, we consider the isothermal pressures [see the terms $\propto v_{tj}$, $j=e$ for electrons and $i$ for ions, in Eqs. \eqref{moment-eq-elctron} and \eqref{moment-eq-ion}].  Also, since the drift velocities of both electrons and ions that are driven by the transverse electric field $E_x\hat{x}$ are in the same direction $\hat{x}$, one can discuss all the physical variables in the $xz$ plane, i.e., in $2D$ geometry. 
\par
Thus, the basic equations for isothermal electrons and ions  read \cite{liting2001} 
\begin{equation}
0=-\frac{e}{m_e} E_z-\frac{v_{te}^2}{n}\frac{\partial n}{\partial z}, \label{moment-eq-elctron}
\end{equation}
\begin{equation}
\frac{\partial n}{\partial t}+\nabla\cdot(n{\bf v}) =0, \label{cont-eq-ion}
\end{equation}
 \begin{equation}
\left(\frac{\partial}{\partial t}+{\bf v}\cdot\nabla\right)v_{z}=\frac{e}{m_i}E_z+\frac{e}{c m_i}v_{x}B_y-\frac{v_{ti}^2}{n}\frac{\partial n}{\partial z},\label{moment-eq-ion} 
\end{equation}
\begin{equation}
v_{x}=\frac{c}{B_0 \omega_{c}}\frac{\partial E_x}{\partial t}, \label{drift-velo}
\end{equation}
where $\nabla=\left(\partial_x,0,\partial_z\right)$,   ${\bf v}=\left(v_x,0,v_z\right)$ is the ion velocity, $m_e~(m_i)$ is the electron (ion) mass, $c$ is the speed of light in vacuum, $\omega_c=eB_0/cm_i$ is the ion-cyclotron frequency, and we have used the quasineutrality condition for the number densities of electrons and ions, i.e., $n_e=n_i=n$ as applicable for long-wavelength ion plasma oscillations.  Furthermore, $v_{tj}=\sqrt{k_BT_j/m_j}$ is the thermal speed of electrons $(j=e)$ or ions $(j=i)$ with $k_B$ denoting the Boltzmann constant  and $T_j$ the thermodynamic temperature.   Also, Eq. \eqref{drift-velo} represents the polarization drift velocity of ions associated with the transverse electric field perturbation $E_z$.  
\par
 The $y$-component of the Faraday's law and the $x$-component of the Amp{\'e}re-Maxwell equation, respectively, are
\begin{equation}
\frac{\partial E_x}{\partial z}-\frac{\partial E_z}{\partial x}=-\frac{1}{c}\frac{\partial B_y}{\partial t}, \label{faraday-law}  
\end{equation}
\begin{equation}
\frac{\partial B_y}{\partial z}=-\frac{4\pi}{c}env_{x} ,\label{ampere-law} 
\end{equation}
where we have neglected the displacement current in Eq. \eqref{ampere-law} due to the low-frequency $(\omega\ll\omega_{c})$ assumption of KAWs.
 \par
 Next, we normalize the physical quantities according to
\begin{equation}
\begin{split}
&n\rightarrow {n}/{n_0},~ B\rightarrow {B_y}/{B_0},~u\rightarrow {v_{z}}/{V_A}, \notag\\
&E_{(x,z)}\rightarrow c{E_{(x,z)} }/{B_0 V_A}, 
(x,z)\rightarrow (x,z) {\omega_c}/{V_A},~t\rightarrow t\omega_c, \label{normalization}
\end{split}
\end{equation}
where $n_0$ is the background number density of electrons and ions and $V_A={B_0}/{\sqrt{4\pi m_in_0}}$  is the Alfv{\'e}n speed. Thus, with the normalizations \eqref{normalization} and eliminating $v_x$ using Eq. \eqref{drift-velo}, we recast the    set of Eqs.      \eqref{moment-eq-elctron} to  \eqref{moment-eq-ion},    \eqref{faraday-law} and \eqref{ampere-law}, respectively,  as
\begin{equation}
\frac{\partial n}{\partial z}=-\alpha^2 n E_z, \label{momentum-eq-electron}
\end{equation}
\begin{equation}
\frac{\partial n}{\partial t}+\frac{\partial}{\partial z}(nu)+\frac{\partial}{\partial x}\left( n \frac{\partial E_x}{\partial t}\right) =0, \label{continuity-eq-ion}
\end{equation}
\begin{equation}
\frac{\partial u}{\partial t}+u\frac{\partial u}{\partial z}+\frac{\partial E_x}{\partial t} \frac{\partial u}{\partial x}=T E_z+B\frac{\partial E_x}{\partial t} ,\label{momentum-eq-ion} 
\end{equation}
\begin{equation}
\frac{\partial E_x}{\partial z}-\frac{\partial E_z}{\partial x}=-\frac{\partial B}{\partial t} , \label{normalized-faraday-law}  
\end{equation}
\begin{equation}
\frac{\partial B}{\partial z}=-n\frac{\partial E_x}{\partial t},\label{normalized-ampere-law} 
\end{equation}
where  $\alpha=V_A/C_s$ is the characteristic speed ratio   with $C_s=\sqrt{{k_B T_e}/{m_i}}$ denoting the ion-acoustic speed and $T=1+T_i/T_e$ which includes the  thermal contributions of electrons and ions. For $T_e\gg T_i$, $T\approx1$.
\par
In order to investigate the amplitude modulation and the nonlinear evolution of  KAW envelopes, we introduce the stretched coordinates by a small scaling parameter $\epsilon~(>0)$ as \cite{misra2011}
\begin{equation}
\xi=\epsilon(x-v_{gx}t),~~  \eta=\epsilon(z-v_{gz}t),~~\tau=\epsilon^2 t, \label{stretch}
\end{equation}
where $v_{gx}$ and  $v_{gz}$ are the group-velocity components of the modulated KAWs along $x$ and $z$ directions. The physical variables are expanded  about their equilibrium values as
\begin{equation}
n=1+\sum_{r=1}^{\infty} \epsilon^r\sum_{l=-\infty}^{\infty}n^{(r)}_l\exp[il({\bf k}\cdot{\bf r}-\omega t)], \label{n-perturbation}
\end{equation}
\begin{equation}
u=\sum_{r=1}^{\infty} \epsilon^r\sum_{l=-\infty}^{\infty}u^{(r)}_l\exp[il({\bf k}\cdot{\bf r}-\omega t)], \label{u-perturbation}
\end{equation}
\begin{equation}
\left(E_x,E_z\right)=\sum_{r=1}^{\infty} \epsilon^r\sum_{l=-\infty}^{\infty} \left(E^{(r)}_{xl},E^{(r)}_{zl}\right)\exp[il({\bf k}\cdot{\bf r}-\omega t)], \label{E-perturbation}
\end{equation}
\begin{equation}
B=1+\sum_{r=1}^{\infty} \epsilon^r\sum_{l=-\infty}^{\infty}B^{(r)}_l\exp[il({\bf k}\cdot{\bf r}-\omega t)], \label{B-perturbation}
\end{equation}
where $n^{(r)}_l$, $u^{(r)}_l$, $E^{(r)}_{xl}$, $E^{(r)}_{zl}$ and $B^{(r)}_l$ satisfy the reality condition $S^{(r)}_{-l}=S^{(r)*}_l$ with asterisk denoting the corresponding complex conjugate   quantity. Also, ${\bf k}=(k\cos\theta,~k\sin\theta)\equiv(k_x,~k_z)$ and the group velocity, $v_g=(v_{gx},~v_{gz})$ with 
 $\theta$ denoting the angle between the wave vector ${\bf k}$ and the $x$-axis.  
\par 
Next, substituting the stretched coordinates, given by Eq. \eqref{stretch} and the expansions, given by Eqs. \eqref{n-perturbation}-\eqref{B-perturbation}, into the normalized equations \eqref{momentum-eq-electron} to \eqref{normalized-ampere-law},  we obtain different  sets of reduced equations for different powers of $\epsilon$ from which one can obtain different harmonic modes. We however, skip the details of the derivations as the method is straightforward and available in the literature \cite{misra2011}. 
\subsection{Linear dispersion law}\label{sec-sub-disp-law}
Equating the coefficients of $\epsilon$ for $n=1,~l=1$, we obtain from Eqs. \eqref{momentum-eq-electron} to \eqref{normalized-ampere-law} the linear dispersion law:  
\begin{equation}
\left(1+\frac{i k_z}{\omega k_x} \right)k_x^2=\left(T-\frac{\omega^2\alpha^2}{k_z^2} \right)\left(\frac{k_z^2}{\omega^2}-1 \right).\label{dispersion-relation} 
\end{equation}
We note that the the imaginary part in Eq. \eqref{dispersion-relation} appears due to the anomalous dissipation of KAWs. In particular, for    $k_x\gg k_z$, applicable to KAWs,  the dispersion equation \eqref{dispersion-relation}  reduces to
\begin{equation}
\omega^2=  \frac{k_z^2}{2\alpha^2}\left(\eta+\sqrt{\eta^2-4\alpha^2T}\right), \label{dispersion-relation-reduced} 
\end{equation}
where $\eta=T+\alpha^2+k_x^2$. Equation \eqref{dispersion-relation-reduced} can be further reduced  in the limit of $T_e\gg T_i$  to that for KAWs as introduced by Hasegawa and Mima, written in terms of the original variables as \cite{hasegawa1976} 
\begin{equation}
\omega^2=k_z^2V_A^2\left(1+k_x^2\rho_i^2\right),
\end{equation} 
where $\rho_i=V_A/\omega_c$ is the ion gyroradius.   From Eq. \eqref{dispersion-relation-reduced}, we find that while the frequency of the KAWs increases with increasing values of $k_x,~k_z$ and $T$,   it decreases with increasing values of $\alpha$. 
\subsection{Compatibility condition}\label{sec-sub-compati-cond}
From the second order reduced equations with $n=2,~l=1$, we obtain from Eqs. \eqref{momentum-eq-electron} to \eqref{normalized-ampere-law}  the following compatibility conditions  for the group velocity components.
\begin{equation}
v_{gx}\equiv\frac{\partial \omega}{\partial k_x}=\frac{2k_x+{ik_z}/{\omega}}{ \Lambda},
\end{equation} \label{vx-component}
with
\begin{equation}
\Lambda=\frac{ik_xk_z}{\omega^2}+\frac{2\omega\alpha^2}{k_z^2}\left(1-\frac{k_z^2}{\omega^2} \right)-\frac{2k_z^2}{\omega^3}\left(T-\frac{\omega^2\alpha^2}{k_z^2} \right), 
\end{equation}
and
\begin{equation}
v_{gz}\equiv\frac{\partial \omega}{\partial k_z}=\frac{\omega}{k_z}.\label{vz-component}
\end{equation}
In the limit of $k_x\gg k_z$, the group velocity component $v_{gx}$ reduces to
\begin{equation}
v_{gx}=\frac{k_xk_z}{\sqrt{2}\alpha}\frac{\sqrt{\eta+\sqrt{\eta^2-4\alpha^2T}}}{\sqrt{\eta^2-4\alpha^2T}}.\label{vx-reduced}
\end{equation}
\subsection{The NLS equation}\label{sec-sub-nls-derivn}
From the equations   for $n=2,~l=0$ and $n=2,~l=2$ we obtain the zeroth and second harmonic modes which involves terms proportional to $\big|B^{(1)}_1\big|^2$ and $\left[B^{(1)}_1 \right]^2$ respectively. These modes so obtained together with the results for the first harmonic modes are then used in the reduced equations for $n=3,~l=1$. Thus, after few steps, we obtain the following 2D nonlocal coupled set of equations for the evolution  of modulated KAW envelopes.
\begin{equation}
\begin{split}
i\frac{\partial B^{(1)}_1}{\partial \tau}&+P_{11}\frac{\partial^2 B^{(1)}_1}{\partial \xi^2}+P_{12}\frac{\partial^2 B^{(1)}_1}{\partial \xi \partial\eta} \\
&+Q_{11}\big|B^{(1)}_1\big|^2 B^{(1)}_1+f^{(2)}_0B^{(1)}_1=0, \label{NLS-equation1}
\end{split}
\end{equation}
\begin{equation}
\begin{split}
\left(v_{gx}\frac{\partial}{\partial\xi} +v_{gz}\frac{\partial}{\partial\eta} \right)^2 f^{(2)}_0 &=D_{1}\frac{\partial^2 \big|B^{(1)}_1\big|^2}{\partial \xi^2}+D_{2}\frac{\partial^2 \big|B^{(1)}_1\big|^2}{\partial \xi \partial\eta}\\
&+D_{3}\frac{\partial^2 \big|B^{(1)}_1\big|^2}{\partial \eta^2},\label{NLS-equation2}
\end{split}
\end{equation}
where $f^{(2)}_0=R_1n^{(2)}_0+R_2u^{(2)}_0$, and the coefficients of   Eqs. \eqref{NLS-equation1} and \eqref{NLS-equation2} are given in Appendix \ref{appendix-a}.  
\par
From Eq. \eqref{vz-component}, we find  that the parallel (to the magnetic field) component of the group velocity  approaches that of the phase velocity of KAWs, implying that the group velocity  dispersion along the external magnetic field ($z-$axis) vanishes, i.e.,  ${\partial^2\omega}/{\partial k^2_z}=0$. Thus, the modulated KAW packet is dispersionless along the $z-$axis.  So, without loss of generality,  we can consider the modulation along the $x-$axis. In this case,  we can neglect the group-velocity component   along the $z$-axis, as well  as the $z$- or $\eta$-dependance of the physical variables for which  Eqs. \eqref{NLS-equation1} and \eqref{NLS-equation2}  reduce to the following NLS equation for KAW envelopes.
\begin{equation}
i\frac{\partial B^{(1)}_1}{\partial \tau}+P\frac{\partial^2 B^{(1)}_1}{\partial \xi^2}+Q|B^{(1)}_1|^2B^{(1)}_1=0, \label{NLS-equation}
\end{equation} 
where $P=P_{11}$ and $Q=Q_{11}+{D_1}/{v_{gx}^2}$.
\section{Analysis of modulational instability}\label{sec-mod-inst}
Before proceeding to the analysis of MI of KAW envelopes, we first note that the group velocity dispersion coefficient $P$ is always real, however, the nonlinear coefficient $Q$ can be complex due to the anomalous dissipation of KAWs. In this case,  as will be shown later, the KAWs are always unstable. However, the dissipation effect can be weak or negligibly small depending on the values of the angle of propagation of the wave vector ${\bf k}$ with the magnetic field ${\bf B}_0$ or $\theta$, and the characteristic speed ratio $\alpha$. Let us scale the smallness of $Q_2$ compared to $Q_1$ as $|Q_2/Q_1|\lesssim o(\epsilon)\sim0.01$, otherwise for $|Q_2/Q_1|>0.01$,  $Q_2$ is said to be comparable to or can even larger than than $Q_1$. Since for kinetic Alfv{\'e}n waves,    $k_x\gg k_z$, we can also assume that $k_z/k_x\sim o(\epsilon)\sim0.01 $.     Thus, one can define two different regimes of $k_x$ for a fixed $k_z~(\ll k_x)$ and $\alpha$: (i) $Q$ is complex with $|Q_2/Q_1|>0.01$ and (ii) $Q$ is real for $|Q_2/Q_1|\lesssim0.01$. In Case (i), the modulated wave is shown to be always unstable for which we calculate the frequency shift and the energy transfer rate, while in Case (ii), we show that the instability sets in for $PQ>0$ (Here, $Q\equiv Q_1$), but still we can have the formation of bright envelope soliton due to the energy localization of KAWs.  
\par
\begin{figure*}[ht]
\centering
\includegraphics[height=2.5in,width=7in]{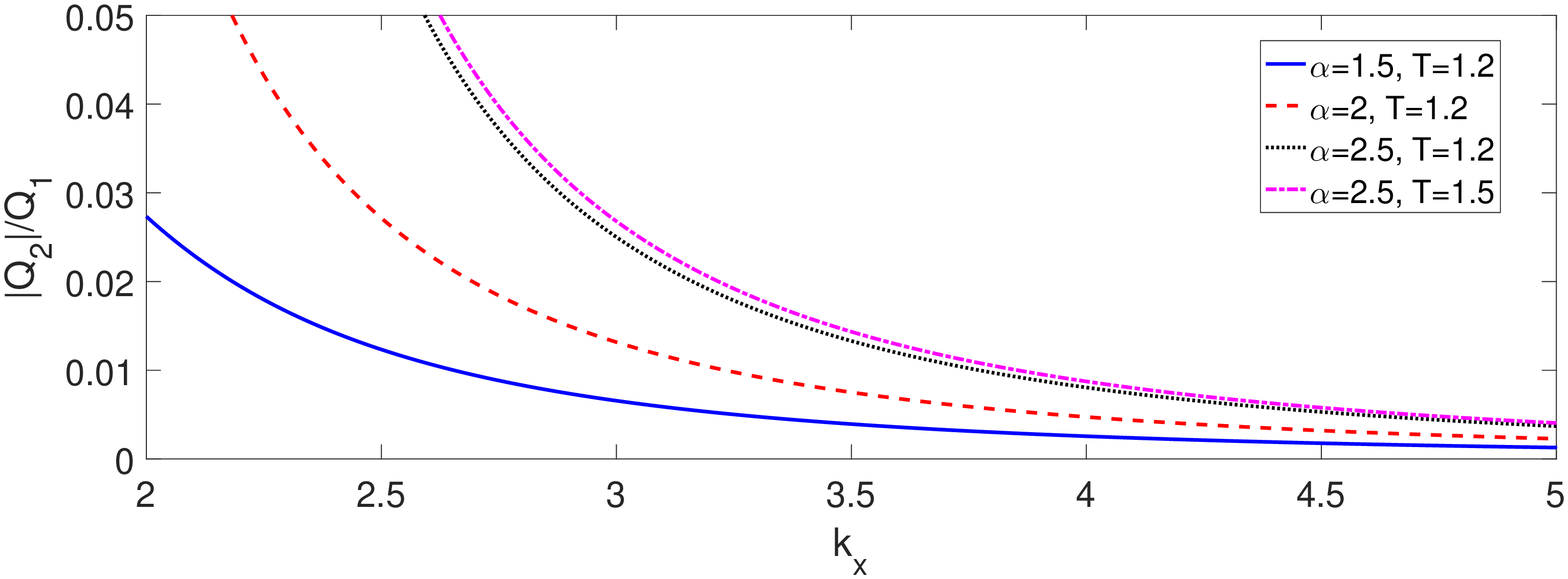}
\caption{ (Color online) $|Q_2|/Q_1$ is plotted against $k_x$ (with a fixed $k_z=0.02$, so that $k_z/k_x\lesssim0.01$) to exhibit the regimes of $k_x$ for  $|Q_2|/Q_1\lesssim0.01$ (i.e., $|Q_2|$ is almost negligible compared to $Q_1$) and  $|Q_2|/Q_1>0.01$   (i.e., either $|Q_2|$ is  comparable to or can be larger than $Q_1$)   for different   values of   $\alpha$  and $T$ as shown in the legend.     }
\label{fig1}
\end{figure*}

Figure \ref{fig1} shows that, given a fixed value of $k_z\sim0.02$ and $\alpha~\sim1/\sqrt{\beta}>1$, where $\beta=8\pi n_0k_B\left(T_e+T_i\right)/B_0^2$, there exit, in fact, two ranges of values of $k_x~(>2)$ (Note here that we have considered $k_x>2$ so that $k_z/k_x\lesssim0.01$ holds), namely $2\leq k_x\lesssim k_{c}$ and $k_x>k_c$.  in   which the relations     $|Q_2/Q_1|>0.01$ and $|Q_2/Q_1|\lesssim0.01$ hold.   However,  as the value of $\alpha$ increases, the  relation $|Q_2/Q_1|>0.01~(\lesssim0.01)$ is satisfied  in  larger (smaller) intervals of $k_x$ (see the dashed and dotted lines).  Similar characteristics are also found with an increase of the ion temperature via the parameter $T$ (compare the dotted and dash-dotted lines).    For example, for a fixed $k_z=0.02$ and $T=1.2$,  as $\alpha$ increases from $\alpha=1.5$, $2$ and $2.5$, the values of $k_c$ increase and so we have    wide ranges of  $k_x$    where $|Q_2/Q_1|>0.01$ holds,   i.e.,  $2\leq k_x \lesssim k_{c1} $,   $2\leq k_x \lesssim k_{c2}$ and     $2 \leq k_x\lesssim k_{c3}$, where $k_{cj}$ is some value of the critical wave number $k_c$ of $k_x$ and  $k_{c1}<k_{c2}<k_{c3}$. For $\alpha=1.5$, $2$ and $2.5$, we have $k_{c1}\sim2.7$, $k_{c2}\sim3.24$ and $k_{c3}\sim3.79$ respectively. The corresponding rages of $k_x$ for which $|Q_2/Q_1|\lesssim0.01$ holds are $k_x>k_{cj}$, $j=1,2,3$, i.e., small ranges  of values of  $k_x$.  Thus,  one can conclude that at a very high value  of $\alpha$, the modulated KAWs become unstable leading to the damping of wave envelope. The corresponding frequency shift may be high and  the transfer of wave energy from high-frequency side bands to low-frequency ones may be slow. On the other hand, at a very low value of $\alpha$ (e.g., the solid line), the modulated KAW  is said to be stable or  unstable depending on the sign of $PQ<0$ or $>0$ $(Q=Q_1)$. Since $P$ is always positive and from Fig. \ref{fig1}, $Q_1>0$, we will have  the  possibility of the onset of growing instability.   However, these features will be discussed in detail shortly.    
\par
In order to investigate the MI, we first (before any modulation) consider   a plane wave solution of Eq. \eqref{NLS-equation} of the form
\begin{equation}
B^{(1)}_1=\rho^{1/2}\exp\left( i\int^\xi \frac{\sigma}{2P}d\xi\right),\label{phi-eq}
\end{equation}
where $\rho$ and $\sigma$ are real functions of $\xi$ and $\sigma$. Substituting   Eq. \eqref{phi-eq} into Eq. \eqref{NLS-equation}, we obtain a   set of equations for the real and imaginary parts. Next, we modulate the wave amplitude and phase by means of the  perturbation expansions as
\begin{eqnarray}
&&\rho=\rho_0+\rho_1 \cos(K\xi-\Omega\tau)+\rho_2 \sin(K\xi-\Omega\tau),\notag\\
&&\sigma=\sigma_1 \cos(K\xi-\Omega\tau) + \sigma_2 \sin(K\xi-\Omega\tau), \label{perturbation-expansion}
\end{eqnarray}
where $\Omega$ and $K$ are, respectively, the wave frequency and the wave number of modulation, and $\rho_0$ is a constant. After linearizing the resulting equations, we obtain from Eq. \eqref{NLS-equation} the following dispersion relation.
\begin{equation}
\Omega^2+2PQ\rho_0K^2 -P^2K^4=0, \label{dispersion}
\end{equation} 
Since $Q$ is, in general, complex, we examine a general solution of Eq. \eqref{dispersion} by letting $\Omega=\Omega_r+i\Gamma$ with $Q=Q_1+iQ_2$ and $\Omega_r,~ \Gamma,~ Q_1$ and $Q_2$ having their real values. Thus, we obtain the following expressions for the frequency shift $\Omega_r$ and the energy transfer rate $\Gamma$.
\begin{eqnarray}
&&\Omega_r=\pm \frac{|K|}{\sqrt{2}}\left[ \left\lbrace (P^2K^2-2\rho_0PQ_1)^2 +(2\rho_0PQ_2)^2\right\rbrace^{1/2}\right. \notag\\
&&\left.+(P^2K^2-2\rho_0PQ_1)\right] ^{1/2},\label{frequency-shift}
\end{eqnarray}
\begin{eqnarray}
&&\Gamma=\mp \frac{|K|}{\sqrt{2}}\left[ \left\lbrace (P^2K^2-2\rho_0PQ_1)^2 +(2\rho_0PQ_2)^2\right\rbrace^{1/2}\right. \notag\\
&&\left.-(P^2K^2-2\rho_0PQ_1)\right] ^{1/2},\label{energy-transfer-rate}
\end{eqnarray}
where we consider the upper (lower) sign for $K>0$ $(K<0)$.
\par
\begin{figure*}[ht]
\centering
\includegraphics[height=2.5in,width=7in]{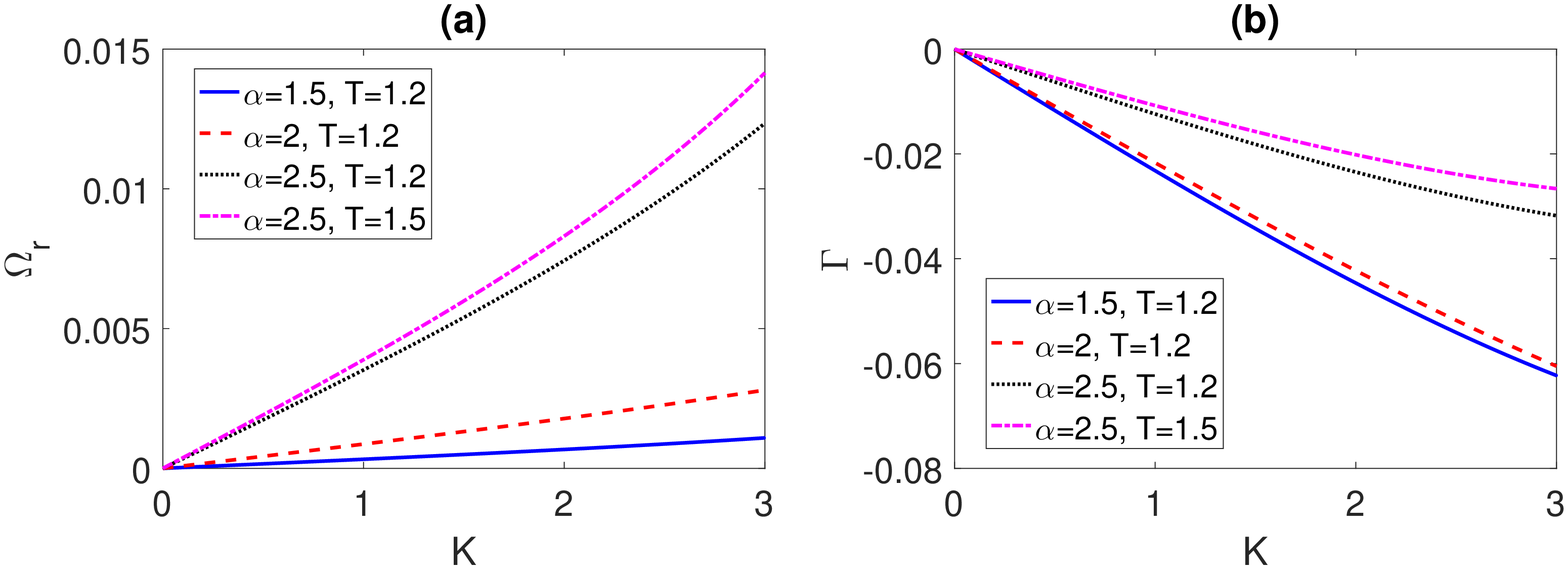}
\caption{ (Color online) The normalized frequency shift $\Omega_r$ [Eq. \eqref{frequency-shift}, left panel (a)] and the energy transfer rate  $\Gamma$ [Eq. \eqref{energy-transfer-rate}, right panel (b)] are plotted against the normalized wave number of modulation $K$ for different values of    $\alpha$ and $T$ as shown in the legends. The other  parameter values are $k_x=2$ and $k_z=0.02$. }
\label{fig2}
\end{figure*}

In what follows, we numerically investigate the characteristics of $\Omega_r$ and $\Gamma$   in the regimes of $k_x$ where  $Q_2/Q_1>0.01$ holds and with the same set of parameters as in Fig. \ref{fig1}. The results are displayed in Fig. \ref{fig2}. We note that  $\Omega_r$ and $\Gamma$ exhibit almost opposite behaviors for  different values of $\alpha$.   It is seen that as     $\alpha$ increases, the frequency shift  $\Omega_r$   increases, but   the energy transfer  rate $|\Gamma|$  decreases. However, the enhancement of $\Omega_r$ or the diminution of  $|\Gamma|$ is significant at higher values of $\alpha>2$.   It means that stronger (weaker) the magnetic field strength (since $\alpha\propto B_0)$, the lower (higher) is the rate of energy transfer from high-frequency side bands to low-frequency ones. We also note that an enhancement of the  electron  and ion thermal energies can also lead to the similar results as for $\alpha$ (compare the dotted and dash-dotted lines).     Thus, it follows that both the magnetic field and the thermal contribution of electrons and ions can influence the  damping of KAWs. 
\par
Next, we consider the case where the anomalous dissipation of KAWs is very weak or almost negligible, i.e.,  in the  regimes of $k_x$ where  $Q_2/Q_1\lesssim0.01$ holds.   In this case, the imaginary part of $Q$ can be neglected, i.e., $Q\approx Q_1$ for which the    dispersion relation \eqref{dispersion} reduces to
\begin{equation}
\Omega^2=P^2K^4\left(1-K_c^2/K^2 \right),\label{real-dispersion} 
\end{equation} 
where  $K_c=\sqrt{{2Q\rho_0}/{P}}$ is the critical wave number of modulation. So, the modulated KAW envelope is said to be stable when $PQ<0$, otherwise, it is unstable for   $PQ>0$ and $K<K_c$. 
 The regions of     $PQ<0$ and  $PQ>0$ are characterized by the wave numbers  $k_x$ and $k_z$  or by the corresponding wave frequency spectra $\omega$ given by the dispersion relation \eqref{dispersion-relation-reduced}.  The dark and bright KAW envelope solitons are thus generated due to the energy localization  and as localized (in space $\sim k^{-1}$ and time $\sim \omega^{-1}$) structures (solutions) of the NLS equation in the cases when $PQ<0$ and $PQ>0$ respectively.
\par
 Since $P$ is always positive and from Fig. \ref{fig1},   $Q_1$ is also seen to be positive in the case of  $Q_2/Q_1\lesssim0.01$,   it follows  that the modulated KAW has a growing instability for $PQ>0$.  The instability growth rate (letting $\Omega\sim i\gamma$) is   given by
\begin{equation}
\gamma=PK^2\left(\frac{K_c^2}{K^2}-1 \right)^{1/2}.\label{growth-rate} 
\end{equation}
Furthermore, the maximum growth rate can be obtained as  $\gamma_{\text{max}}=\rho_0|Q|$. Note that though, in this case, the wave amplitude grows,    we will see shortly  that some physical processes intervene  to stop the growth rate with cut-offs at finite values of the wave number of modulation $K$, leading to the formation of bright envelope solitons.  An analytic form  of the bright   envelope soliton  can be obtained from the NLS equation \eqref{NLS-equation} as \cite{misra2014} $B_1^{(1)}=\sqrt{\cal{B}}\exp(i\theta)$, where  
\begin{equation}
\begin{split}
&{\cal B}={\cal B}_{\text{br}}{\text{sech}}^2\left(\frac{\xi-\tilde{v}_{gx}\tau}{W_{\text{br}}}\right),\\
&\theta=\frac{\tilde{v}_{gx}}{2P}\left[\xi-\left(\frac{\tilde{v}_{gx}}{2}-\frac{PQ{\cal B}_{\text{br}}^2}{\tilde{v}_{gx}}\right)\tau\right]\equiv \frac{\tilde{v}_{gx}}{2P}\left(\xi-\tilde{v}_{px}\tau\right). \label{bright-soliton}
\end{split}
\end{equation}    
Here, $W_{\text{br}}=\sqrt{2P/Q{\cal B}_{\text{br}}}$  and ${\cal B}_{\text{br}}$ is the constant wave amplitude. We note from  $\xi-\tilde{v}_{gx}\tau=\epsilon\left(x-v_{gx}t\right)-\tilde{v}_{gx}\epsilon^2t=\epsilon\left[x-\left(v_{gx}+\epsilon \tilde{v}_{gx}\right)t\right]=\epsilon\left(x-v'_{gx}\right)t$ that $\tilde{v}_{gx}$ represents the excess velocity of the wave envelope over the linear group velocity $v_{gx}$ and $\tilde{v}_{px}$ represents the phase velocity relative to the linear group velocity.  
Equation \eqref{bright-soliton} also shows that the bright envelope oscillates at a frequency $\Omega_0=PQ{\cal B}_{\text{br}}^2$ at rest.
\par
\begin{figure*}[ht]
\centering
\includegraphics[height=2.5in,width=7in]{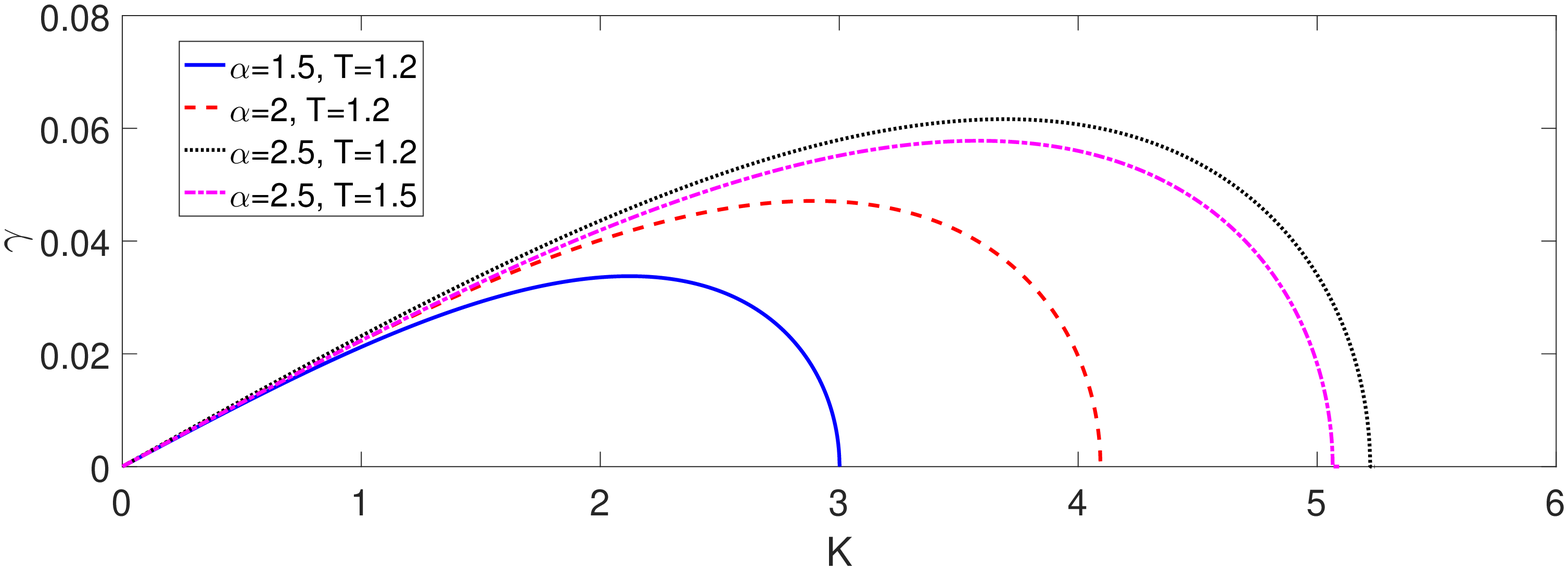}
\caption{ (Color online) The MI growth rate, given by Eq. \eqref{growth-rate}, is plotted   with the dimensionless wave number of modulation $K$ for different values of the parameters $\alpha$ and $T$ as in the legend. The other parameter values are    $k_x=4$ and $k_z=0.02$.  It is seen that the growth rate becomes higher (lower) at higher values of $\alpha$ $(T)$ with cut-offs at higher (lower) values of $K$. }
\label{fig3}
\end{figure*}

Figure \ref{fig3} displays the MI growth rate, given by Eq. \eqref{growth-rate} for the same set of parameters as  in Fig. \ref{fig1}. We find that for a fixed value of $T$, as the value of $\alpha$ increases (see the solid and dashed lines), the growth rate increases having a cutoff at higher wave number of modulation $K$. However, in contrast, as the value of $T$ increases with a fixed value of $\alpha$ (see the dotted and dash-dotted lines), the growth rate decreases with a cutoff  at a lower value  of $K$.    
 \section{Conclusion}\label{sec-conclusion}
 We have investigated the nonlinear amplitude modulation of KAWs in an intermediate low-$\beta$ magnetoplasma. Starting from a set of 2D fluid equations  coupled to the Maxwell's equations and using the standard reductive perturbation technique, we have derived a   coupled set of nonlinear partial differential equations   which govern the slow modulation of KAW packets. The modulational instability analysis is then carried out  by  a NLS equation derived from the   coupled set of equations. It is found that the parameters, namely the angle of propagation of KAWs with the magnetic field (or the parallel and perpendicular components $k_z$ and $k_x$ of the wave vector ${\bf k}$), the enhanced temperature ratio $T\equiv1+T_i/T_e$,  and the characteristic speed ratio $\alpha\equiv V_A/C_s\propto B_0$ play significant roles on the   instability of KAW envelopes.  It is found that for some  fixed values of $k_z~(\ll k_x)$, $\alpha~(>1)$ and $T$, there are two different regimes of $k_x$ in one of which the anomalous dissipation   of KAWs has considerable effect and in other regime it can be negligible.   In fact, the anomalous dissipation can be effective for a wide range of values of $k_x$ at some higher values of $\alpha$. This means that increasing the magnetic field strength favors the instability of KAWs.   When the dissipation effect is not negligible, the modulated KAW is shown to be always unstable. In the process of three-wave interactions,  the corresponding  frequency shift and the energy transfer rate (due to wave damping) from high-frequency side bands to low-frequency ones are calculated for a fixed pump wave amplitude. It is seen that while the energy transfer rate $|\Gamma|$ decreases, the frequency $\Omega_r$ is always up shifted and gets enhanced with increasing    values of $\alpha$. On the other hand, when the dissipation  is no longer effective or the imaginary part of $Q$ becomes negligible compared to its real part, the   KAW  is always found to be unstable (growing instability) for $PQ>0$ and $K<K_c$.  In this case, a higher value of the parameter $\alpha$  $(T)$ is found to enhance (diminish) the growth rate of instability and   give rise cut-offs  at higher (lower) values of $K$.  
 \par 
 It may be noted that the evolution of KAW envelopes, which are associated with  the coupling of transverse  Alfv{\'e}n waves and longitudinal ion-acoustic waves, may be relevant to the coherent and incoherent interactions of waves in intermediate low-$\beta$ plasmas. Here,  the coherent interaction may be  responsible  for a frequency up-shift associated with each wave number  without any energy exchange   between the wave numbers, and the incoherent interaction may be responsible for spectral decay in which the redistribution of energy takes place in the wave number space \cite{sturrock1957}.
 \par We also mention that the theory of kinetic Alfv{\'e}n wave envelopes in intermediate plasmas has not been considered before. However, the excitation of large amplitude intermediate shocks and solitons from kinetic Alfv{\'e}n waves has been reported in Ref. \onlinecite{liting2001} by Liting \textit{et al}. They have shown that the Alfv{\'e}n solitons or shocks exist depending on  whether $k_x$ or $k_z$ and the characteristic ratio $\alpha=V_A/C_s$  are   below or above some critical values. In our work, we have also shown that similar critical values of $k_x$ and $\alpha$ exist for the evolution of Alfv{\'e}n wave envelopes and the occurrence  of wave damping. Due to lack of experimental and observational data in interplanetary space, we could not compare our results with the existing ones.   
 \par
 Furthermore, at    a  point of the solar wind  where the solar wind speed equals the Alfv{\'e}n speed, the kinetic Alfv{\'e}n waves are known to be excited, and they  propagate along the magnetic field lines in different branches. These branches of the kinetic Alfv{\'e}n wave  propagating towards the sun are carried out by the solar wind outward from the sun. Thus, it is more likely that these branches may be localized  and evolve into kinetic Alfv{\'e}n wave envelopes or get damped due to anomalous dissipation in interplanetary space plasma environments. These KAWs may also propagate as bright   envelope solitons when the anomalous dissipation effect is  negligibly small. 
 \par
  For the present study of  kinetic Alfv{\'e}n waves we have neglected the electron inertia.  However,   such effect can be significant in the evolution of inertial Alfv{\'e}n waves \cite{rinawa2015}. These waves can be studied by the similar approach  as in the present work. Also, since KAWs exhibit growing instability and the growth rate of instability becomes high in strong magnetic fields, there may be a possibility of the excitation of rogue waves \cite{misra2014} or kinetic Alfv{\'e}n rogons in space plasmas, which may be a project for future work. 
   \acknowledgments{The authors gratefully acknowledge some useful comments of the Reviewers    which helped improve the manuscript in the present form. This work was supported by UGC-SAP (DRS, Phase III) with Sanction  order No.  F.510/3/DRS-III/2015(SAPI),  and UGC-MRP with F. No.
43-539/2014 (SR) and FD Diary No. 3668.}
 \appendix
 \section{The coefficients of   Eqs. \eqref{NLS-equation1} and \eqref{NLS-equation2} } \label{appendix-a}
 \begin{widetext}
 \begin{equation}
P_{11}=\frac{k_z^2}{4\alpha^2\omega\sqrt{\beta^2-4\alpha^2T}}\left[(\beta+\sqrt{\beta^2-4\alpha^2T}) (k_x^2+\sqrt{\beta^2-4\alpha^2T}) 
-\frac{8k_x^2T\alpha^2}{\sqrt{\beta^2-4\alpha^2T}} \right]\equiv \frac{1}{2}\frac{\partial^2\omega}{\partial k_x^2},\label{P_1} 
\end{equation} 

 \begin{eqnarray}
P_{12}= &&\left[ \left\lbrace \omega \alpha^2 k_z v_{gx}\left(\frac{k_z^2}{\omega^2}+1 \right)+k_z(Tk_z^2-\omega^2\alpha^2)\left(\frac{k_z^2}{\omega^3}+\frac{k_z^2}{\omega^2}-\frac{1}{k_x} \right)  
+\frac{Tk_z^3}{\omega}\left(\frac{k_z^2}{\omega^2}+1 \right)\right\rbrace v_{gx}\right. \notag\\ &&\left.-2\alpha^2 k_z^2 v_{gx}  -\frac{2Tk_z^4}{\omega^2}-\frac{k_z^2}{\omega k_x} 
\left(Tk_z^2-\omega^2\alpha^2\right)\left(\frac{k_z}{\omega}+2 \right)\right. \notag\\
&&\left. +\omega^2\alpha^2 \left(\frac{k_z^2}{\omega^2}-1 \right)v_{gx}+i\frac{k_xk_z^3}{\omega}\left( 1+v_{gx}\frac{k_z}{\omega}\right) \right]  
\big/\left[-\omega k_z^3\left\lbrace\frac{ik_xk_z}{\omega^2}-\frac{2}{\omega^3 k_z^2}(Tk_z^4-\omega^4\alpha^2) \right\rbrace \right]\equiv \frac{1}{2}\frac{\partial^2\omega}{\partial k_xk_z},\label{P_{12}} 
\end{eqnarray}
\begin{eqnarray}
Q_{11}=&&\left[\omega\alpha^2\mu_n k_x k_z^2+ T\omega k_z^3 \left(\frac{k_z^2}{\omega^2}-1 \right)\mu_n  -\omega k_x^2 k_z^3\mu_n+\alpha^2\omega^2k_z^2 \left(\frac{k_z^2}{\omega^2}-1 \right)\mu_u +Tk_z^4\left(\frac{k_z^2}{\omega^2}-1 \right)\mu_u -2 k_x^2 k_z^4\mu_u \right.\notag\\
&&\left. -2k_x^2k_z^4\mu_x +k_z(\omega^2 \alpha^2 -Tk_z^2)\mu_n -i\alpha^4 \omega^3\left(\frac{k_z^2}{\omega^2}-1 \right)\mu_z+ik_xk_z^4\mu_n -2i\alpha^2\omega^3 k_xk_z^2\left(\frac{k_z^2}{\omega^2}-1 \right)\mu_x+ i\frac{k_xk_z^5}{\omega}\mu_u\right.\notag\\
&&\left. +i2\omega k_xk_z^3T\left(\frac{k_z^2}{\omega^2}-1 \right)\mu_x+ i2\omega k_xk_z^3\mu_x -ik_xk_z^4\mu_B +i\frac{2\omega^2\alpha^2}{k_xk_z}(\omega^2\alpha^2 -Tk_z^2)\left(\frac{k_z^2}{\omega^2}-1 \right)\mu_x\right] \notag\\
&& \big/\left[ \left(\frac{k_z^2}{\omega^2}-1 \right)\left(\omega^2 \alpha^2k_z -\frac{iT}{k_x} \right)+2k_z(Tk_z^2-\omega^2\alpha^2)+k_x^2k_z^3 \right],\label{Q} 
\end{eqnarray}
With
\begin{eqnarray}
\mu_x= && \left[ \frac{k_z^3}{\omega k_x}\left\lbrace 3 k_x^2 -2\left(\frac{k_z^2}{\omega^2}-1 \right) \left(T+\frac{\omega^2 \alpha^2}{k_z^2} \right)\right\rbrace  +ik_z^2\left(1-k_x^2 +\frac{2k_z^2}{\omega^2}+\left(\frac{k_z^2}{\omega^2}-1 \right) \left\lbrace T\left(1+\frac{\omega^2\alpha^2}{2k_x^2k_z^2}\right) + \frac{2\omega^2\alpha^2}{k_z^2}\left(1-\frac{\omega^2\alpha^2}{4k_x^2k_z^2} \right)  \right\rbrace  \right)\right]  \notag\\
&& / \omega\left\lbrace(i-2)k_z-3\omega k_x \right\rbrace ,\label{mu-x}
\end{eqnarray}

\begin{eqnarray}
\mu_z= &&\frac{\omega^2}{k_x k_z}\left(\frac{k_z^2}{\omega^2}-1 \right)\left[ \mu_x-\frac{i\omega^2\alpha^2}{2 k_x} \right] , \label{mu-z}
\end{eqnarray}
\begin{eqnarray}
\mu_n=&&\frac{\alpha^2}{2k_z^2}\left[i\mu_z k_z- \frac{\omega^2\alpha^2}{k_x^2}\left(\frac{k_z^2}{\omega^2}-1 \right)^2 \right], \label{mu-n} 
\end{eqnarray}
\begin{eqnarray}
 \mu_u=&& \frac{1}{k_z}\left[\omega \mu_n +i 2\omega k_x \mu_x +\alpha^2 \left(\frac{k_z^2}{\omega^2}-1 \right)\left\lbrace-\frac{ik_z}{k_x} +\frac{T\omega}{k_x^2}\left(\frac{k_z^2}{\omega^2}-1 \right) -\omega \right\rbrace  \right], \label{mu-u} 
 \end{eqnarray}
 \begin{equation}
 \mu_B=(k_z \mu_x-k_x\mu_z)/\omega, \label{mu-B}
 \end{equation}
\begin{eqnarray}
 D_1=-\frac{2Tk_z}{k_x}\left(\frac{k_z^2}{\omega^2}-1 \right)R_2v_{gx}+ \left\lbrace 1+2T\left(\frac{k_z^2}{\omega^2}-1 \right)\right\rbrace \frac{k_z}{\omega}v_{gx}^2-R_1\frac{\omega \alpha^2}{k_x}\left(\frac{k_z^2}{\omega^2}-1 \right)v_{gx},\label{D_1}
 \end{eqnarray}
  \begin{eqnarray}
 D_2=&&v_{gx}\left[\lambda_u^2 R_2-R_1\frac{k_z}{\omega}\left\lbrace 1+2T\left(\frac{k_z^2}{\omega^2}-1 \right) \right\rbrace  \right]-\frac{2Tk_z}{k_x}R_2\left(\frac{k_z^2}{\omega^2}-1 \right)v_{gz} +2\frac{k_z}{\omega}\left\lbrace 1+2T\left(\frac{k_z^2}{\omega^2}-1 \right) \right\rbrace v_{gx}v_{gz}\notag\\
 &&-R_1\frac{\omega \alpha^2}{k_x}\left(\frac{k_z^2}{\omega^2}-1 \right)v_{gz}+\frac{2T\omega \alpha^2}{k_x^2k_z}R_1\left(\frac{k_z^2}{\omega^2}-1 \right)^2 v_{gx}+R_1 \frac{2Tk_z}{k_x}\left(\frac{k_z^2}{\omega^2}-1 \right),\label{D_2}
 \end{eqnarray}
  \begin{eqnarray}
 D_3=&&v_{gz}\left[\lambda_u^2 R_2-R_1\frac{k_z}{\omega}\left\lbrace 1+2T\left(\frac{k_z^2}{\omega^2}-1 \right) \right\rbrace  \right]+\frac{k_z}{\omega}v_{gz}^2\left\lbrace 1+2T\left(\frac{k_z^2}{\omega^2}-1 \right) \right\rbrace +R_1\lambda_u^2 \notag\\
 &&+R_1\frac{2T\omega \alpha^2}{k_x^2k_z}\left(\frac{k_z^2}{\omega^2}-1 \right)^2 v_{gz},\label{D_3}
 \end{eqnarray}
with
\begin{eqnarray}
R_1=-\left(T-\frac{\omega^2\alpha^2}{k_z^2}\right)\big/\left[\frac{ik_xk_z}{\omega^2}-\frac{2}{\omega^3k_z^2}(Tk_z^4-\omega^4\alpha^2)\right],\label{R_1}
\end{eqnarray}
 \begin{eqnarray}
R_2=\frac{i}{\omega}\left[2\omega^2\alpha^2\left(\frac{k_z^2}{\omega^2}-1 \right)+k_xk_z^2\right]\big/\left[\frac{ik_xk_z}{\omega^2}-\frac{2}{\omega^3k_z^2}(Tk_z^4-\omega^4\alpha^2)\right].\label{R_2}
\end{eqnarray}
\end{widetext}

\end{document}